\documentclass[fleqn,usenatbib,useAMS]{mnras}
\usepackage{mathptmx}
\usepackage[T1]{fontenc}
\usepackage{ae,aecompl}

\usepackage{graphicx}	% Including figure files
\usepackage{amsmath}	% Advanced maths commands
\usepackage{amssymb}	% Extra maths symbols

\newcommand{\kev}{keV}

\newcommand{\bat}{\textit{Swift}-BAT}

\newcommand{\etal}{et al.}

\newcommand{\clf}{$\Psi(L|M_{\mathrm{h}})$}
\newcommand{\hmf}{$n(M_{\mathrm{h}})$}
\newcommand{\mh}{$M_{\mathrm{h}}$}

\title[A New Era of Black Hole Demographics -- II. The CLF of Type 2 AGNs]{Clustering,
  Cosmology and a New Era of Black Hole Demographics -- II. The
  Conditional Luminosity Functions of Type 2 and Type 1 Active Galactic Nuclei}

\author[D. R. Ballantyne]{
D. R. Ballantyne\thanks{E-mail: david.ballantyne@physics.gatech.edu}
\\
Center for Relativistic Astrophysics, School of Physics, Georgia
  Institute of Technology, 837 State Street, Atlanta, GA 30332-0430, USA\\
}

\date{Accepted XXX. Received YYY; in original form ZZZ}

\pubyear{2016}

\begin{document}
\label{firstpage}
\pagerange{\pageref{firstpage}--\pageref{lastpage}}
\maketitle

\begin{abstract}
The orientation-based unification model of active galactic nuclei (AGNs)
posits that the principle difference between obscured (Type 2) and
unobscured (Type 1) AGNs is the line-of-sight into the central
engine. If this model is correct than there should be no difference in
many of the properties of AGN host galaxies (e.g., the mass of the surrounding
dark matter haloes). However, recent clustering analyses of Type 1 and Type 2
AGNs have provided some evidence for a difference in the halo mass, in
conflict with the orientation-based unified model. In this work, a
method to compute the Conditional Luminosity Function (CLF) of Type 2
and Type 1 AGNs is presented. The CLF allows many fundamental halo
properties to be computed as a function of AGN luminosity, which we
apply to the question of the host halo
masses of Type 1 and 2 AGNs. By making use of the total AGN CLF, the
Type 1 X-ray luminosity function, and the luminosity-dependent Type 2
AGN fraction, the CLFs of Type 1 and 2 AGNs are calculated at $z
\approx 0$ and $0.9$. At both $z$, there is no statistically
significant difference in the mean halo mass of Type 2 and 1 AGNs at
any luminosity. There is marginal evidence that Type 1 AGNs may have
larger halo masses than Type 2s, which would be consistent with an
evolutionary picture where quasars are initially obscured and then
subsequently reveal themselves as Type 1s. As the Type 1 lifetime is
longer, the host halo will increase somewhat in mass during the Type 1
phase. The CLF technique will be a powerful way to study the
properties of many AGNs subsets (e.g., radio-loud, Compton-thick) as
future wide-area X-ray and optical surveys substantially increase our
ability to place AGNs in their cosmological context.
\end{abstract}


%%%%%%%%%%%%%%%%%%%%%%%%%%%%%%%%%%%%%%%%%%%%%%%%%%

%%%%%%%%%%%%%%%%% BODY OF PAPER %%%%%%%%%%%%%%%%%%

\section{Introduction}
\label{sect:intro}
The unified model of active galactic nuclei (AGN) has been a central pillar
of AGN phenomenology for nearly three decades. At its core, the model
simply states that the central engines and environs of all AGNs are
essentially the same, but, because of an axisymmetric absorbing
structure (the `torus'), objects may have different observational
properties depending on the viewing angle to the AGN \citep[e.g.,][]{anton93,up95,netzer15}. In this
way, AGNs of similar luminosities which are observed to suffer from significant line-of-sight
absorption (i.e., Type 2 AGNs) differ from the unobscured, Type 1 AGNs
solely due to a line-of-sight that passes through the torus. There
is substantial evidence that this orientation-based unification model
is broadly applicable in local Seyfert galaxies where the nuclear regions can
be studied in detail at infrared wavelengths \citep[e.g.,][]{burt13,honig13,trist14}. However,
it is clear that this model is too simple and requires detailed
adjustment to account for the wide range of obscuration properties
seen at different $z$ and luminosities \citep[e.g.,][]{merloni14}. For example, the fact
that the fraction of obscured AGNs appears to increase with $z$
indicates that the obscuring region is connected to the overall
evolution of the host galaxy \citep{laf05,bem06,tu06,has08,ueda14,buch15}. As the torus is
situated at scales of $1$--$100$~pc from the central engine, it
provides a bridge between the galactic and black hole environments,
and therefore contains information on the processes responsible for
fueling the nuclear activity. 

An alternative to the orientation-based unified model is one based on
the evolution of the sources. In this model, an event triggers
luminous AGN
activity that is initially obscured due to the large amount of gas
funneled into the nuclear environment. Feedback from the AGN
eventually clears out much of the absorbing gas and the AGN shines as
a Type~1 source before fading due to the dwindling fuel supply \citep[e.g.,][]{san88,kh00}. This evolutionary scenario has been shown to occur in
numerical simulations of major merger-triggered AGNs \citep[e.g.,][]{hernq89,dsh05,hopk05}. With suitable
adjustment of parameters, this model can explain many aspects of AGN
phenomenology \citep[e.g.,][]{hopk06,hopk08}. Despite this success, observations at $z \ga 1$ only
uncover evidence of merger triggering at the highest AGN
luminosities \citep[e.g.,][]{koc12,trei12,glik15,delmoro16}. Determining the redshift and luminosity range in which the
orientation-based unification model holds therefore provides crucial
information on the triggering mechanisms of these AGNs. Regions in the
redshift-luminosity plane where the unified model fails are more
likely to be subject to the drastic fueling events needed for the
evolutionary scenario. 

Since the orientation-based unified model states that the main difference
between Type 2 and Type 1 AGNs is the view through the parsec-scale
obscuring torus, the galactic environment of AGN host galaxies
should be independent of the level of nuclear obscuration. One way to
quantitatively test this prediction is to measure the clustering
properties (such as the average AGN bias, $\bar{b}_{\mathrm{A}}$, or
mean dark matter halo mass, $\langle M_{\mathrm{h}} \rangle$), of both
Type 1 and Type 2 AGNs. In the orientation-based unification model,
these quantities should be similar for both populations. Performing
this test is challenging because large samples of both types of AGNs
need to be identified over wide areas. However, recently several
groups have utilized large \textit{Spitzer} and \textit{WISE}
IR-selected samples of quasars to investigate the clustering
properties of Type 1 and 2 AGNs at $z \sim 1$. After separating the
two classes by an optical/IR colour cut, some authors found that obscured
AGNs were more biased and thus populated more massive haloes than
unobscured ones \citep*[e.g.,][]{hick11,don14,dipomp14,dhm15}, with a difference in halo mass ranging from a
factor of $2$ to $10$. The statistical significance of the difference
also varies greatly from study to study, and indeed other groups
report no difference between the two types
\citep{geach13,mendez16}. Interestingly, \citet{allev11,allev14} used X-ray selected AGNs in the COSMOS field and found the
opposite tendency at $z \sim 2$ and $3$; that is, Type 1 AGNs were more
biased and inhabited more massive haloes than obscured AGNs (see also
\citealt{cap10}). Direct comparison between the X-ray and IR results
is problematic due to the strong selection effects that arise from
constructing each sample \citep{mendez16}. Nevertheless,
there are now several results indicating that Type 1 and 2 AGNs may inhabit
different environments, in contrast to the orientation-dependent
unification picture.

These recent observational results probe only certain AGN luminosities, but if the
lower-luminosity Seyfert galaxies are likely triggered by different processes
than quasars \citep[e.g.,][]{db12} the nature of AGN obscuration could change with
luminosity. While performing clustering analyses of Type 1 and 2 AGNs
in narrow luminosity bins is not possible with current data, future
wide-field surveys by \textit{eROSITA}, \textit{Euclid},
\textit{WFIRST} and LSST with allow this experiment. In preparation
for this new era of AGN demographics Ballantyne (2016; hereafter
Paper I), presented a method to compute the Conditional Luminosity
Function (CLF) of AGNs using X-ray survey data. The CLF allows the
computation of multiple statistics of how AGNs populate dark matter
haloes as a function of luminosity. Paper I computed the CLF at both
$z \approx 0$ and $\approx 0.9$ and showed that the mean halo mass of
AGNs does increase with luminosity at both epochs, indicating how the
fueling of AGNs differs with luminosity and $z$.

In this paper, the X-ray CLF methodology developed in Paper I is modified so
that CLFs of Type 1 and Type 2 AGNs can be individually
constrained at $z \approx 0$ and $0.9$ (Sects.~\ref{sect:CLF} and~\ref{sect:apply}). The CLF-derived statistics on the
host halo properties of Type 2 and Type 1 AGNs are described in
Sect.~\ref{sect:results}. The implications of these results on the AGN
unification model and triggering physics are discussed and summarized in
Section~\ref{sect:discuss}. As in Paper I, a WMAP9 $\Lambda$CDM
cosmology is assumed in this work: $h=0.7$, $\Omega_m=0.279$,
$\Omega_{\Lambda}=0.721$ and $\sigma_8=0.821$ \citep{hinshaw13}.

\section{The Conditional Luminosity Function of Type 2 and Type 1 AGNs}
\label{sect:CLF}
This section first gives a brief overview of the key definitions and
equations behind the AGN CLF methodology presented in Paper I. This method
closely follows the CLF model for galaxies first described by
\citet{yang03} and \citet{vym03}. The description is first focused on
the entire AGN population, and then Sect.~\ref{sub:CLFhowto} presents
how to break down the total CLF to ones for the Type 2 and Type 1 AGN
population.

\subsection{Brief Review of CLF Definitions}
\label{sub:CLFintro}
The AGN CLF, \clf, is defined at a specific $z$ so that the XLF is
obtained when the CLF is integrated over all halo masses; i.e.,
\begin{equation}
\label{eq:clf}
\phi(L)=\int_{0}^{\infty} \Psi(L|M_{\mathrm{h}}) n(M_{\mathrm{h}})
dM_{\mathrm{h}},
\end{equation}
where $\phi(L)=d\Phi/d\log L$ is the AGN XLF in a specific energy band
(here, $2$--$10$~\kev), and \hmf\ is the dark matter halo mass
function. A crucial aspect of the CLF is that it provides a
statistical description of how AGNs of different luminosities are
distributed in halos of various masses. Therefore, once constrained
the CLF can be used to calculate several statistical measures of the
AGN population, such as the mean number of AGNs with luminosities
between $L_1$ and $L_2$ as a function of halo mass, \mh:
\begin{equation}
\label{eq:avgN}
\left < N(M_{\mathrm{h}}) \right > = \int_{L_1}^{L_2}
\Psi(L|M_{\mathrm{h}}) dL.
\end{equation}
From this, the average \mh\ hosting an AGN with a luminosity in this 
interval of $L$ is \citep{yang03}
\begin{equation}
\label{eq:avgM}
\left <M_{\mathrm{h}}(L) \right >  = {1 \over \phi(L)} \int_{0}^{\infty} M_{\mathrm{h}}
  \left < N(M_{\mathrm{h}}) \right >  n(M_{\mathrm{h}})
  dM_{\mathrm{h}}.
\end{equation}
The CLF can also be used to estimate the lifetime of AGNs by following
the argument of \citet{mw01}. Paper I showed that the AGN lifetime as
a function of luminosity can be estimated as 
\begin{equation}
\label{eq:time}
\tau_{\mathrm{AGN}}(L) = (\Psi(L|\left <M_{\mathrm{h}}(L) \right >) dL) \tau_{\mathrm{Hubble}},
\end{equation}
where $\tau_{\mathrm{Hubble}}$ is the Hubble time at the redshift of
interest. Indeed, all of the above expressions are evaluated at a
specific $z$, but the $z$-dependence is not included for the sake of
clarity. 

As it is a statistical description of the AGN population in the
cosmological environment, the CLF must be parameterized and
constrained by observational data instead of being derived from a
fundamental theory. Paper I showed that the following
parameterization of the CLF was appropriate for AGNs at $z \approx 0$ and
$\approx 0.9$:
\begin{equation}
\label{eq:clfform}
\Psi(L|M_{\mathrm{h}}) = \left( {M_{\mathrm{h}} \over M_{\ast}}
\right )^{a} e^{-M_{\mathrm{h}}/M_{\mathrm{cut}}} f(L)
\end{equation}
where
\begin{equation}
\label{eq:mcut}
M_{\mathrm{cut}}=\left( {L \over L_{\ast}} \right)^c M_{N},
\end{equation}
and
\begin{equation}
\label{eq:brokenpower}
f(L)= \left \{ \begin{array}{ll}
\left( {L \over
  L_{\ast}} \right)^{-0.96} & \mbox{if $L < L_{\ast}$} \\
\left( {L \over
  L_{\ast}} \right)^{-\beta} & \mbox{otherwise.}
\end{array} \right.
\end{equation}
This form for the CLF has 6 free parameters ($a$, $M_{\ast}$, $\beta$,
$L_{\ast}$, $c$, and $M_{N}$) and was chosen to give a broken
power-law XLF and a cutoff power-law $\left < N(M_{\mathrm{h}}) \right
>$, both consistent with observations \citep[e.g.,][]{ueda14,lea15}.

Following \citet{yang03}, Paper I used the observed AGN XLF plus
measurements of the AGN correlation length at different luminosities,
$r_0(L)$, to constrain the CLF parameters. The correlation lengths,
defined as the radius where the AGN two-point correlation function,
$\xi_{\mathrm{AA}}(r)$ is unity, are typically large enough that the
`2-halo' contribution dominates the correlation function. The `2-halo' component describes the clustering of AGNs hosted in separate dark
matter haloes, and is related to the clustering of the haloes themselves,
\begin{equation}
\label{eq:xi2h}
\xi_{\mathrm{AA}}^{2h}(r) \approx \bar{b}_A^2 \xi_{\mathrm{dm}}^{2h},
\end{equation}
where $\xi_{\mathrm{dm}}^{2h}$ is the `two-halo' contribution to the total
dark matter correlation function (defined as $\xi_{\mathrm{dm}}^{2h}=\xi_{\mathrm{dm}}-\xi_{\mathrm{dm}}^{1h}$), and 
\begin{equation}
\label{eq:bbar}
\bar{b}_A={\int_{0}^{\infty} n(M_{\mathrm{h}}) \left <
    N(M_{\mathrm{h}}) \right > b(M_{\mathrm{h}}) dM_{\mathrm{h}} \over
  \int_{0}^{\infty} n(M_{\mathrm{h}}) \left <
    N(M_{\mathrm{h}}) \right > dM_{\mathrm{h}}}
\end{equation}
is the mean AGN bias ($b(M_{\mathrm{h}})$ is the bias of dark
matter haloes to the dark matter distribution). As
$\xi_{\mathrm{dm}}^{2h}$ and $b(M_{\mathrm{h}})$ can be computed from
a cosmological model, the CLF allows
$\bar{b}_A$ and $\xi_{\mathrm{AA}}^{2h}(r)$ to be calculated for
different luminosity ranges. As before, these quantities are also
functions of $z$. The \citet{tink10} fitting formulas for \hmf\ and
$b(M_{\mathrm{h}})$ are used in the calculations, and further
information on the cosmological details needed in the CLF modeling can
be found in Appendix A of Paper I.

\subsection{Determining the Type 2 and Type 1 CLFs}
\label{sub:CLFhowto}
In Paper I, the AGN CLF was constrained at $z \approx 0$ and $0.9$ and
the above statistical measurements of the total AGN population were
examined at these two redshifts. \citet{vym03} showed how the total
CLF of galaxies could be separated into ones for early-type and
late-type galaxies. Here, the method of \citet{vym03} is adapted in
order to determine the CLFs for Type 2 and Type 1 AGNs. Once known,
the equations from Sect.~\ref{sub:CLFintro} can be used to calculate
statistics for each set of AGNs.

We begin by noting that the AGN CLF can be decomposed into the sum of
the CLF for Type 2 AGNs, $\Psi(L|M_{\mathrm{h}})_2$, and the CLF for
Type 1 AGNs, $\Psi(L|M_{\mathrm{h}})_1$; i.e.,
\begin{equation}
\label{eq:clfsum}
\Psi(L|M_{\mathrm{h}})=\Psi(L|M_{\mathrm{h}})_1+\Psi(L|M_{\mathrm{h}})_2.
\end{equation}
In the case considered here and in Paper I, the above statement is
true for Compton thin AGNs, as the number of Compton thick sources are
negligible in the samples used to constrain the CLFs.

By defining $f_2(L,M_{\mathrm{h}})$ as the fraction of Type 2 AGNs
with luminosity $L$ in haloes of mass \mh, the Type 2 and Type 1 CLFs
can be written as
\begin{equation}
\label{eq:ty2clf}
\Psi(L|M_{\mathrm{h}})_2
dL=f_2(L,M_{\mathrm{h}})\Psi(L|M_{\mathrm{h}})dL
\end{equation}
and
\begin{equation}
\label{eq:ty1clf}
\Psi(L|M_{\mathrm{h}})_1
dL=(1-f_2(L,M_{\mathrm{h}}))\Psi(L|M_{\mathrm{h}})dL.
\end{equation}
Integrating $f_2(L,M_{\mathrm{h}})$ over mass or luminosity gives the
mean AGN Type 2 fraction as a function of luminosity or mass,
respectively; i.e., 
\begin{equation}
\label{eq:ty2vsL}
\bar{f}_2(L) = {\int f_2(L,M_{\mathrm{h}}) \Psi(L|M_{\mathrm{h}})
  n(M_{\mathrm{h}}) dM \over \int \Psi(L|M_{\mathrm{h}})
  n(M_{\mathrm{h}}) dM }
\end{equation}
and
\begin{equation}
\label{eq:ty2vsM}
\bar{f}_2(M_{\mathrm{h}}) = {\int f_2(L,M_{\mathrm{h}})
  \Psi(L|M_{\mathrm{h}}) dL. \over \int \Psi(L|M_{\mathrm{h}}) dL}
\end{equation}
The left hand side of Eq.~\ref{eq:ty2vsL} can be measured
observationally, but the exact dependence of the Type 2 fraction with
luminosity is much debated \citep*{burlon11,merloni14,sck15}.

To determine $f_2(L,M_{\mathrm{h}})$, we again follow \citet{vym03}
and assume it is separable; i.e.,
$f_2(L,M_{\mathrm{h}})=g(L)h(M_{\mathrm{h}})$ (this assumption is
checked in Appendix~\ref{app:self}). With this
assumption $g(L)$ can be written as
\begin{equation}
\label{eq:gofL}
g(L) = \bar{f}_2(L) {\int \Psi(L|M_{\mathrm{h}}) n(M_{\mathrm{h}}) dM \over
  \int \Psi(L|M_{\mathrm{h}}) h(M_{\mathrm{h}}) n(M_{\mathrm{h}}) dM}
\end{equation}
Thus, since $\bar{f}_2(L)$ is observed for AGNs, if $h(M_{\mathrm{h}})$ can
be constrained, $g(L)$ and hence $f_2(L,M_{\mathrm{h}})$,
$\Psi(L|M_{\mathrm{h}})_2$, and $\Psi(L|M_{\mathrm{h}})_1$ all can be
determined. Once the individual CLFs are known, they can be used to
calculate all the standard CLF-derived statistics described in
Sect.~\ref{sub:CLFintro} for both Type 2 and Type 1 AGNs.

\section{Application to AGNs at $z \approx 0$ and $0.9$}
\label{sect:apply}
The decomposition of the AGN CLF into the sum of the Type 2 and Type 1
CLFs requires first knowing the total CLF at the redshift of
interest. In Paper I, the total AGN CLF was determined at $z \approx
0$ and $z \approx 0.9$. Now, these CLFs will be used along with
estimates of $f_2(L)$ to derive individual CLFs for Type 2 and 1 AGNs
at both redshifts.

As in Paper I, an AGN XLF will be used to constrain the CLFs. Here, we
make use of the Type 1 XLFs determined by \citet*{has05} using
AGNs selected in the $0.5$--$2$~keV band (the XLFs are
converted to the $2$--$10$~\kev\ band employed here using the same
procedure and spectral model as \citet{ball14}). The Type 1 AGNs used
to measure the XLFs were
identified by the presence of broad Balmer lines in
the optical spectrum or X-ray hardness ratios. There is generally
good agreement between the optical and X-ray definitions of Type 1 and
2 AGNs \citep[e.g.,][]{has08,merloni14,burt16}; thus, we make
use of X-ray determined $f_2(L)$ estimates at $z \approx 0$
\citep{burlon11} and $0.9$ \citep{merloni14}. The \citet{burlon11} study of \bat-detected
AGNs is hard X-ray selected and shows the least bias for detecting
Compton-thin AGNs (but see \citealt{sck15}). \citet{burlon11} found that the fraction
of obscured AGNs (defined as those with column densities $>
10^{22}$~cm$^{-2}$) could be described as
\begin{equation}
\label{eq:f2burlon}
f_2^{z\approx 0}(L)=0.8e^{-L/10^{43.7}}+0.2(1-e^{-L/10^{43.7}}),
\end{equation}
where the $L$ is defined in the $15$--$55$~keV band (we
assume the same form for the $2$--$10$~\kev\ band, as the changes in
luminosities between the two bands are relatively small). 
There is some evidence that the Type 2 fraction increases with $z$,
especially at $\log (L/\mathrm{erg\ s^{-1}}) \ga 44$ \citep[e.g.,][]{ueda14}. To capture this effect,
Eq.~\ref{eq:f2burlon} is revised so that the Type 2 fraction roughly
follows the $z \approx 0.9$ estimates of \citet{merloni14}:
\begin{equation}
\label{eq:f2merloni}
f_2^{z\approx 0.9}(L)=0.75e^{-L/10^{44.2}}+0.5(1-e^{-L/10^{44.2}}).
\end{equation}

The procedure to determine $\Psi(L|M_{\mathrm{h}})_2$, and
$\Psi(L|M_{\mathrm{h}})_1$ at each $z$ is as follows. For a
specific parameterization of $h(M_{\mathrm{h}})$, the total AGN CLF
from Paper I is
used along with the appropriate $f_2(L)$ to compute $g(L)$
(Eq.~\ref{eq:gofL}). Then
$f_2(L,M_{\mathrm{h}})=g(L)h(M_{\mathrm{h}})$ is determined while
enforcing the condition that $f_2(L,M_{\mathrm{h}}) \leq 1$ for all
$L$ and $M_{\mathrm{h}}$. The Type~1 AGN CLF,
$\Psi(L|M_{\mathrm{h}})_1$, is then calculated (Eq.~\ref{eq:ty1clf}),
and from that the Type~1 XLF can be computed (Eq.~\ref{eq:clf}) and
compared to the observed datapoints. As in Paper I, a $\chi^2$ is
computed and the \citet{metro53} algorithm is used to adjust any
$h(M_{\mathrm{h}})$ parameters in order to minimize the $\chi^2$
calculated from Type 1 XLF. There are 19 datapoints defining the $z \approx
0$ Type 1 XLF and 11 for the one at $z \approx 0.9$. However, the $z
\approx 0.9$ Type 1 XLF contains AGNs between $z=0.8$ and $1.6$, and
the two highest luminosity points (which would be most heavily
influenced by AGNs at higher $z$) predicted space densities larger
than the total \citet{ueda14} XLF at $z \approx 0.9$. Therefore, we
omit these two points when determining the CLFs.  

After many trials it was found that the simple form
$h(M_{\mathrm{h}})=e^{-M_{\mathrm{h}}/M_{\ast}}$, where $M_{\ast}$ is
the only parameter, provides the best
fit to the XLFs for these forms of $f_2(L)$. More
complex forms of $h(M_{\mathrm{h}})$ did not significantly improve the
fits to the XLFs, but may be required for different forms of $f_2(L)$
or for different XLF shapes. The resulting values of $M_{\ast}$ and
the $\chi^2$ obtained are shown in Table~\ref{table:res}.
\begin{table}
\centering
\caption{Results of fitting the \citet{has05} Type 1 XLF using the total AGN
  CLF from Paper I, $f_2(L)$ (either Eq.~\ref{eq:f2burlon}
  or~\ref{eq:f2merloni}), and assuming
  $h(M_{\mathrm{h}})=e^{-M_{\mathrm{h}}/M_{\ast}}$. The error-bars are calculated using a
  $\Delta \chi^2=2.71$ criterion (i.e., a 90\% confidence level for
  the parameter of interest). Due to the small error-bars and some
  wiggles in the published XLF, the error-bars of the $z \approx 0$
  XLF are increased by $40$\% to lower the minimum reduced $\chi^2$ to
  $1.2$ (there is no change in the fit
parameter). No upper-bound on $M_{\ast}$ was obtained at $z \approx
0$ (the maximum $\Delta \chi^2$ in that direction was $1.28$).}
\label{table:res}
\begin{tabular}{lcc}
\hline
Redshift & $\chi^2/$dof & $\log (M_{\ast}/$M$_{\odot})$\\
\hline
$0$ & $21.4/18$ & $13.97_{-0.31}^{+?}$ \\
$0.9$ & $9.8/8$ & $14.17_{-0.19}^{+0.45}$ \\
\hline
\end{tabular}
\end{table}
Initially, the minimum reduced $\chi^2$ at $z \approx 0$ was $>2$,
largely due to the small error-bars and wiggles in the published XLF. Increasing the error-bars by $40$\% lowered the
minimum reduced $\chi^2$ to $1.2$ with no change in the fit
parameter. At $z \approx 0.9$, the reduced $\chi^2$ is $1.2$, so no
changes were needed to the published error-bars. The best value of
$\log M_{\ast}$ is larger at $z \approx 0.9$ than at $z
\approx 0$, although they are consistent within the
errorbars (defined as the $90$\% confidence
level for one degree of freedom; i.e., a $\Delta \chi^2=2.71$
criterion). However, an upper error-bar could not be found at that confidence
level at $z \approx 0$ (the maximum $\Delta \chi^2$ obtained in that direction is
$1.28$). This is because the CLF model associates high mass haloes
with high luminosity AGNs (Paper I), and the fraction of Type 2 AGNs at quasar
luminosities at $z \approx 0$ is small enough that there is little
sensitivity to how the Type 2 AGNs are distributed in haloes with
$M_{\mathrm{h}} > M_{\ast}$.

\begin{figure*}
\includegraphics[width=0.48\textwidth]{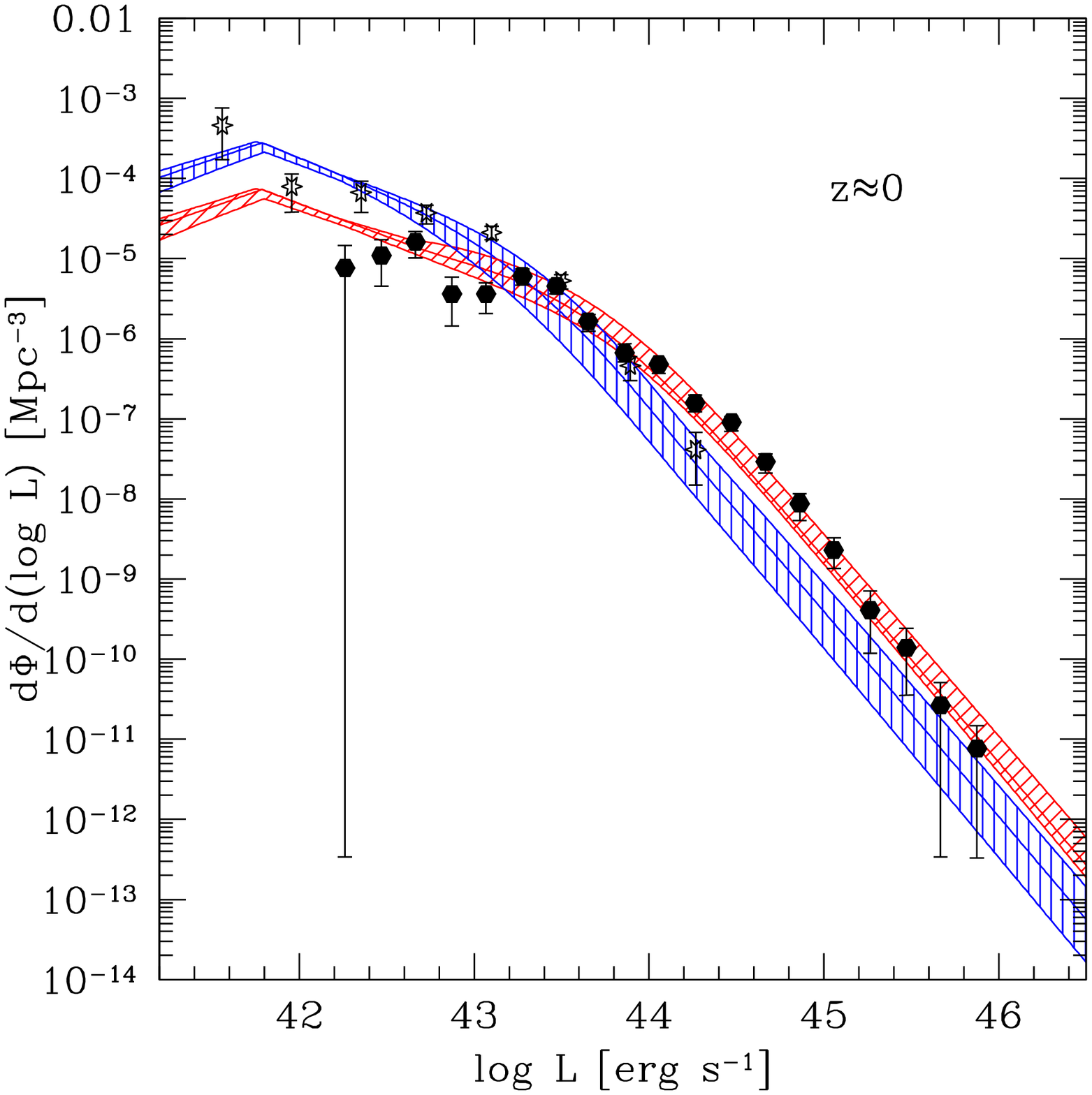}
\includegraphics[width=0.48\textwidth]{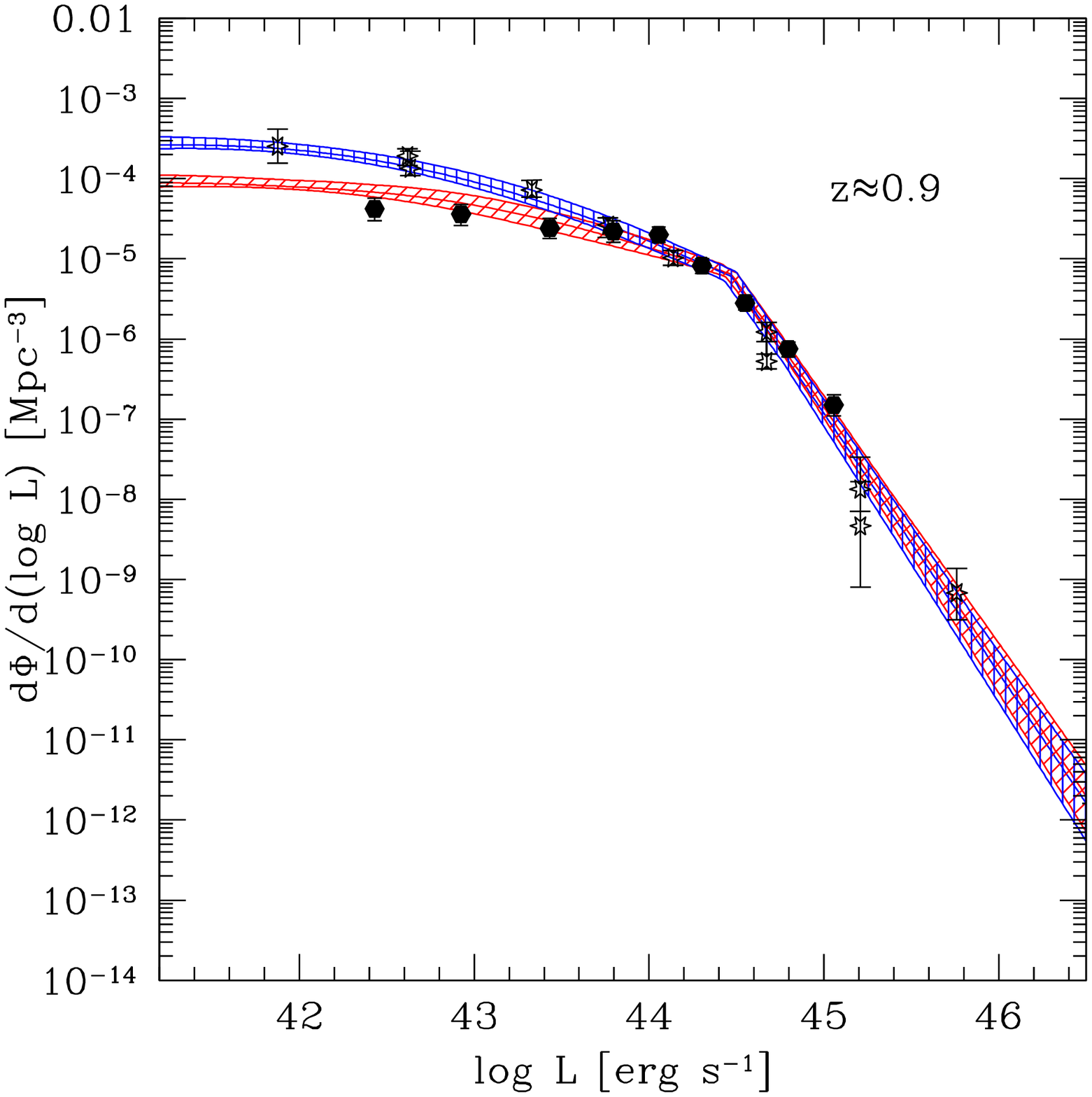}
\caption{(Left) The predicted $2$--$10$~keV Type 2 (blue lines and
  hatched region) and Type 1
  (red lines and hatched region) AGN XLFs computed at $z \approx 0$ from the best fit
  CLFs (Eqs.~\ref{eq:ty2clf} and \ref{eq:ty1clf}). The solid points
  are from the \citet{has05} measurement of the Type~1 AGN XLF translated to the
  $2$--$10$~keV band using the procedure of \citet{ball14}. The open
  stars are the Type~2 XLF measured by \citet{burlon11} using \bat\
  (also converted to the $2$--$10$~keV band). The CLFs were
  constrained by only fitting the Type~1 XLF, but clearly do an
  excellent job describing the Type~2 XLF at this $z$. The hatched
  regions denote the 95\% confidence levels of the predicted XLFs (see
text for details). (Right) As in the left-hand panel, but now showing
the predicted XLFs at $z \approx 0.9$. The solid points are the Type~1
AGN XLF as measured by \citet{has05} at $z = 0.8-1.6$, after omitting the two
highest luminosity points. The open stars plot the $z=0.9$ Type 2 AGN XLF
calculated from the \citet{ueda14} XLF data using their
$f_2(L)$ prescription (their Eq.~3). The CLF-derived Type 2 XLF
remains consistent with the \citet{ueda14} XLF.}
\label{fig:xlf}
\end{figure*}

Figure~\ref{fig:xlf} plots the predicted Type~2 and
Type~1 XLFs at both redshifts derived from the best-fit CLF models.
The \citet{has05} Type~1 XLF that was used to constrain the CLF is show as the
solid points in both panels. In addition, the $z \approx 0$ Type~2 XLF
measured by \citet{burlon11} using \bat\ is shown as the open stars in
the left-hand panel of Fig.~\ref{fig:xlf}. These data were not used to
help constrain the Type 2 CLF as they are not independent from the $f_2(L)$
measurement, but it is clear that the CLF-derived Type~2 XLF provides an
excellent description of the data at this $z$. The right-hand panel
also plots (as open stars) the \citet{ueda14} $z=0.9$ Type~2 XLF
calculated by multiplying their total XLF data with their prescription for $f_2(L)$ (their
Eq. 3). The CLF-derived Type~2 XLF at $z\approx 0.9$ is consistent
with the \citet{ueda14} data despite the different assumed forms of
$f_2(L)$ and our use of the \citet{has05} data to constrain the
XLFs. This agreement indicates that our
procedure produces a $\Psi(L|M_{\mathrm{h}})_2$ and
$\Psi(L|M_{\mathrm{h}})_1$ that is self-consistent with the underlying
total CLF computed in Paper I.

The hatched regions in Fig.~\ref{fig:xlf} are the 95\% confidence
regions of the derived XLFs, calculated using a similar Monte-Carlo
method described in Paper I. Briefly, starting from its best-fit value $M_{\ast}$ is
randomly varied, and a $\chi^2$ is calculated. If this $\chi^2$ is
within the 95\% confidence region for one degree of freedom (a $\Delta
\chi^2 \leq 3.84$ criterion) the value of $M_{\ast}$ is
stored. After 300 random changes, $M_{\ast}$ is reset to the best-fit
value, and the process restarts until a total of 3000 changes have
been considered. To take into account the uncertainty in the total
AGN CLF, the above procedure is performed 10 times: once with the
best-fitting CLFs from Paper I, and nine other times with the six
CLF parameters randomly varied within their 90\% errorbars (see Table 1 in
Paper I). After completion, a total of 3958 (5841) CLF models for $z
\approx 0 (0.9)$ within the confidence region are stored. These are then searched to find the
minimum and maximum values of the quantity of interest as a function
of luminosity or halo mass. The confidence regions are then defined as
the space between the maxima and minima and are shown in all future
plots. 

\section{Results}
\label{sect:results}
In this section, the CLF-derived statistics are used to determine if
there is a significant difference in dark matter halo properties of
obscured and unobscured AGNs.

\begin{figure*}
\includegraphics[angle=-90,width=0.48\textwidth]{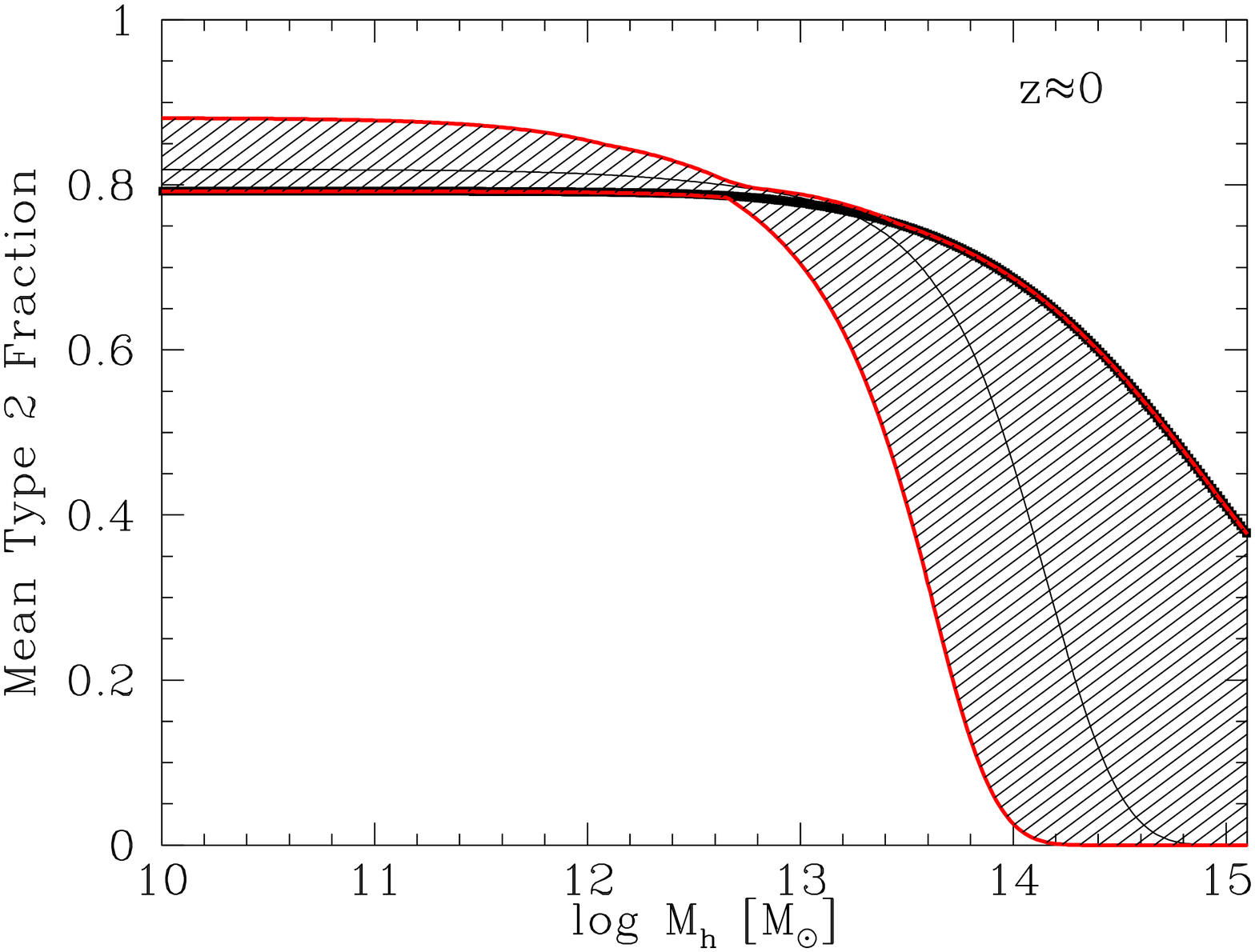}
\includegraphics[angle=-90,width=0.48\textwidth]{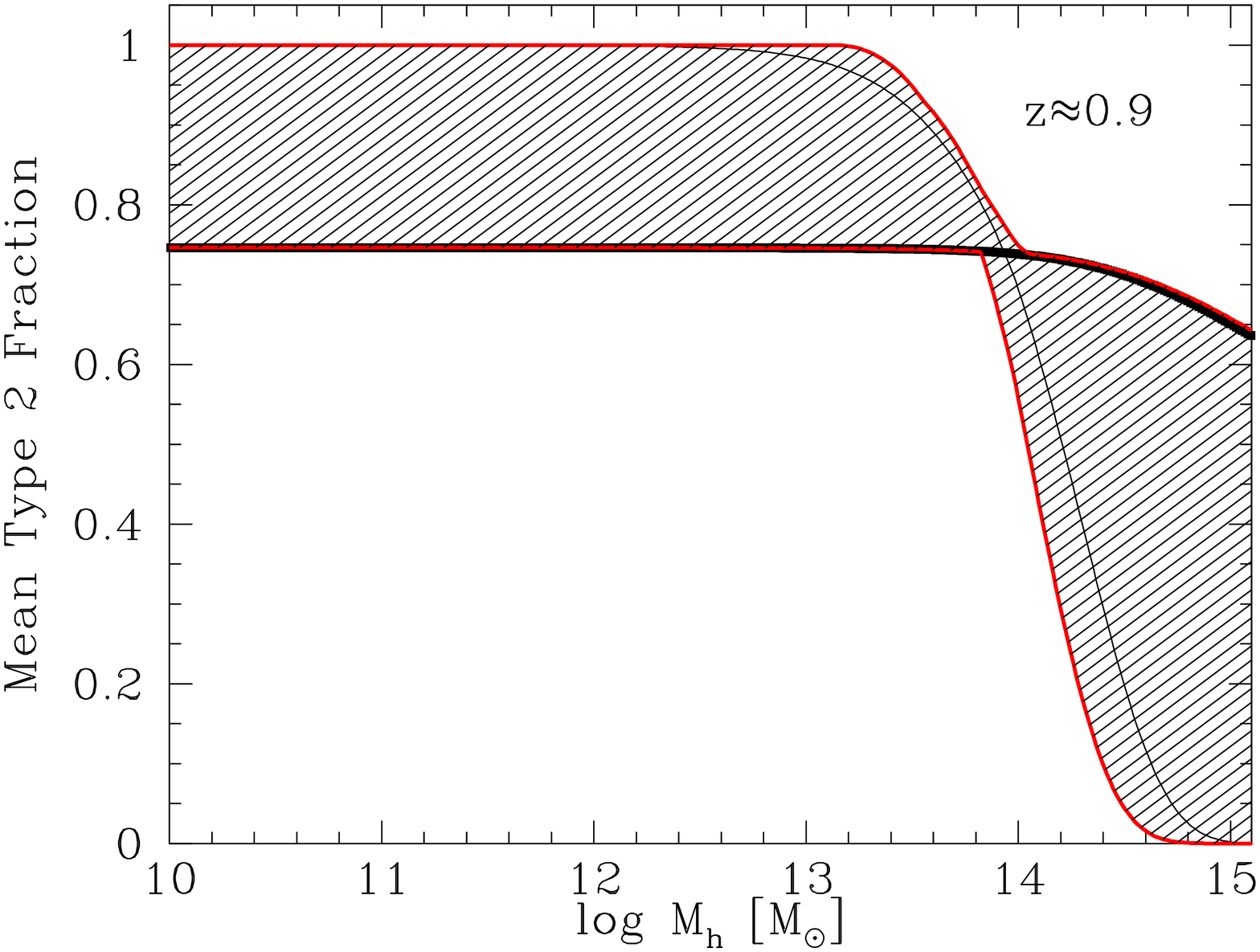}
\caption{(Left) The thin solid line plots the mean Type~2 AGN fraction as a function of
  halo mass at $z \approx 0$ as derived from the AGN CLF
  (Eq.~\ref{eq:ty2vsM}). The hatched region indicates the 95\%
  confidence region. The thick black line is the expected variation in the
  mean Type 2 fraction when there is \emph{no difference} in the halo or
  clustering properties between Type 1 and Type 2 AGN. As this line is
consistent with the boundary of the 95\% confidence region (red lines), there is
only marginal evidence for a difference in the mean Type 2 fraction at $z \approx
0$. (Right) As in the left-hand panel, but at $z \approx 0.9$. Similar
to the situation at lower redshift, the predicted variation in the
mean Type
2 fraction with \mh\ is not significantly different from the one
expected when obscured and unobscured AGNs have identical clustering
properties.}
\label{fig:f2vsM}
\end{figure*}

Before using the individual Type 2 and Type 1 AGN CLFs to derive
population statistics, it is illuminating to use Eq.~\ref{eq:ty2vsM} to
simply calculate how the mean fraction of Type~2 AGNs depends on
$M_{\mathrm{h}}$. The thin solid line in Fig.~\ref{fig:f2vsM} shows that
at both redshifts
the mean Type~2 fraction seems to depend significantly on halo mass when
$\log (M_{\mathrm{h}}/\mathrm{M_{\odot}}) \ga 13$.
However, it is important to recognize that some dependence of the mean
Type~2 fraction on \mh\ is expected because of two observational
facts: first, the Type~2 fraction drops at high AGN luminosities
(Eqs.~\ref{eq:f2burlon} \&~\ref{eq:f2merloni}), and, second, the correlation length
increases with luminosity indicating that high luminosity AGNs are
found in higher mass haloes (\citealt{cap10,kou13}; Paper I). Thus, a
smaller Type 2 fraction is expected at high halo masses regardless of
the clustering properties of the two populations. To illustrate this,
the thick black lines in both panels show $\bar{f}_2(M_{\mathrm{h}})$ in the
situation where the bias, \mh\ and other clustering properties of Type
1 and 2 AGNs are identical (operationally, this is computed by setting
$M_{\ast}$ to a very large value in $h(M_{\mathrm{h}})$). These curves
overlap with the boundary of the 95\% confidence regions of $\bar{f}_2(M_{\mathrm{h}})$
computed from our CLF procedure (red lines). Therefore, it appears that at both $z
\approx 0$ and $0.9$, there is only very weak evidence for a
difference in the host halo properties of Type 1 and 2 AGNs.

\begin{figure*}
\centering
\includegraphics[angle=-90,width=0.48\textwidth]{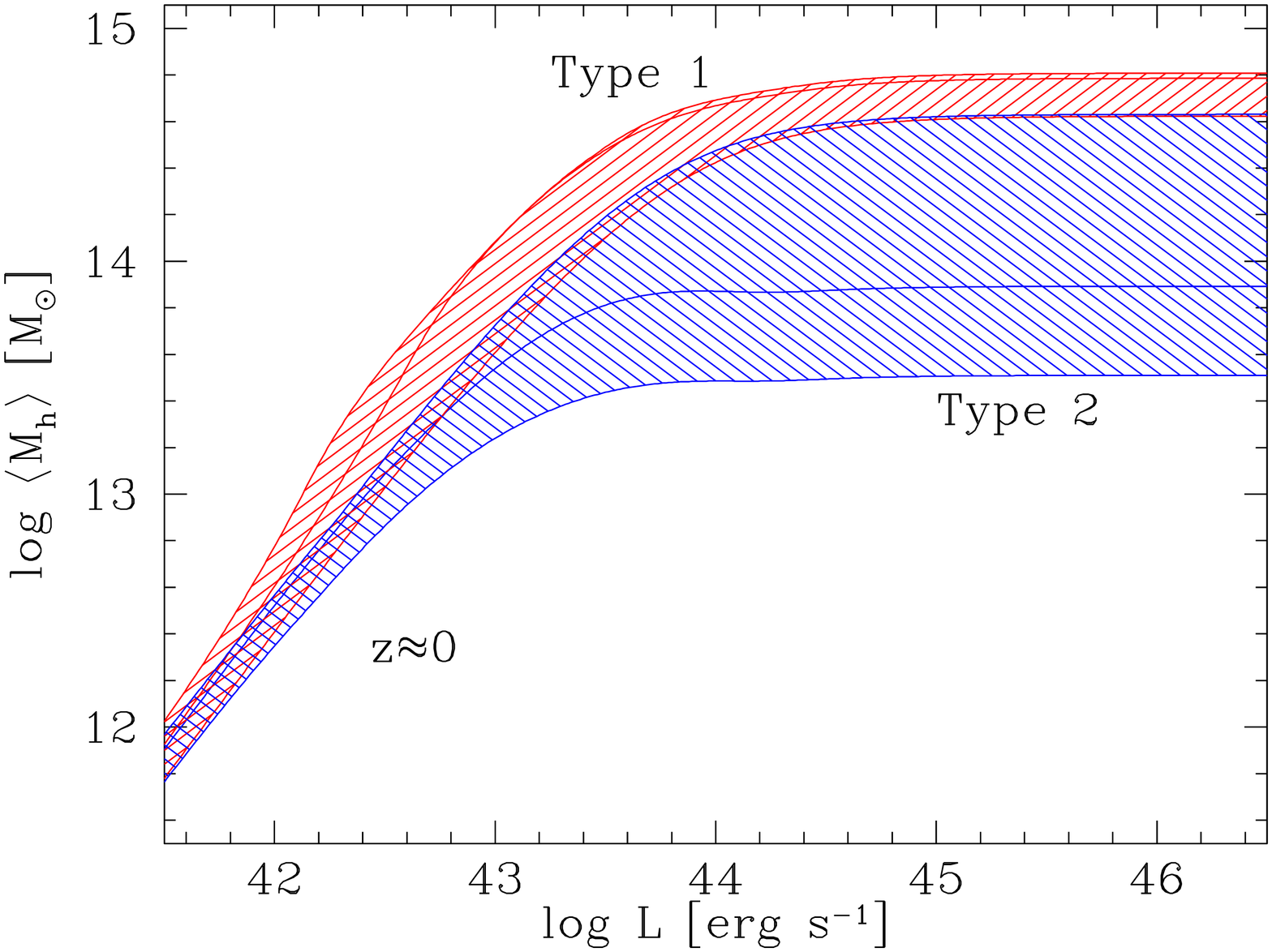}
\includegraphics[angle=-90,width=0.48\textwidth]{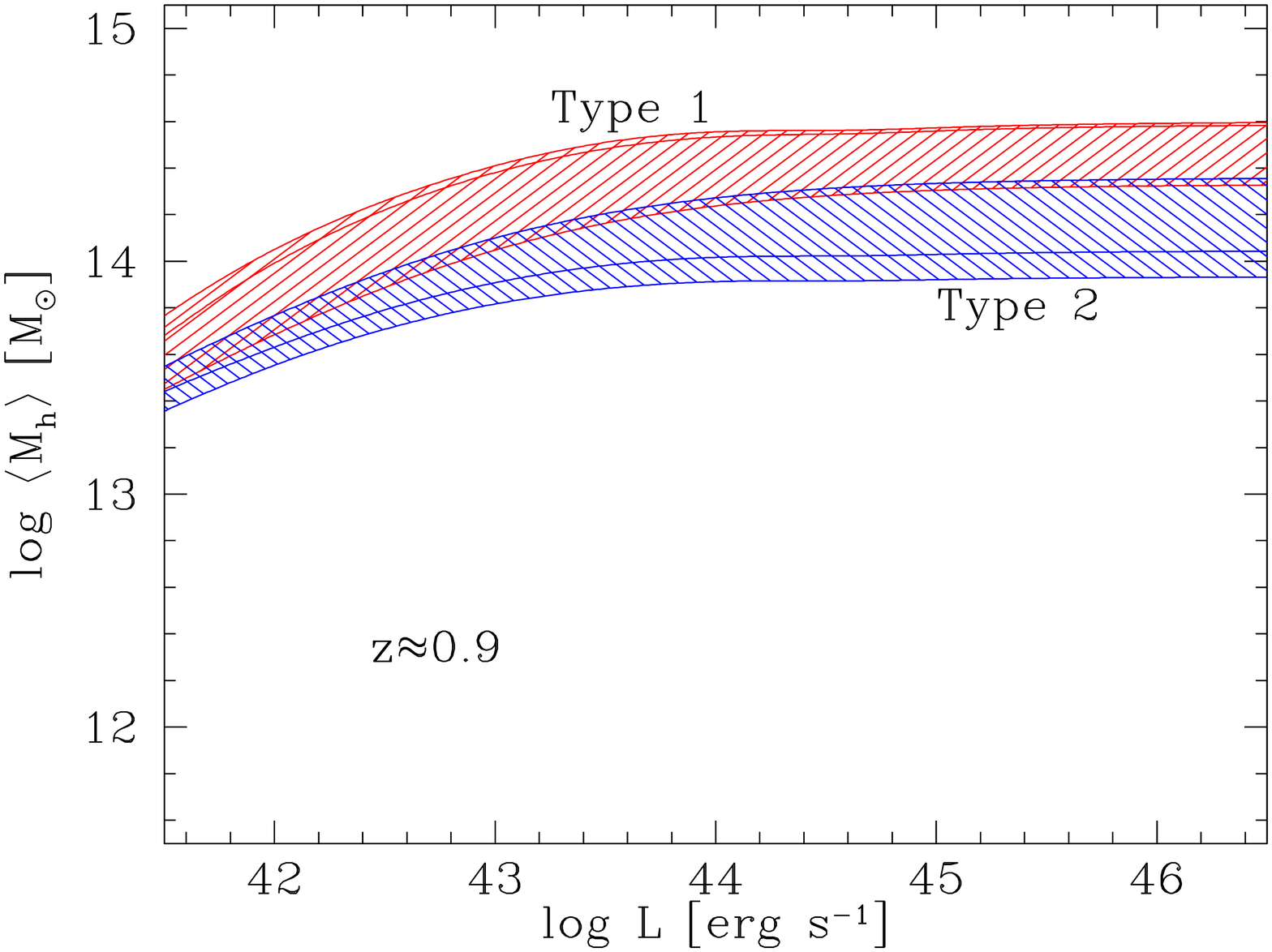}
\caption{The CLF-derived Type 2 (blue) and Type 1 (red) AGN mean halo mass,
$\langle M_{\mathrm{h}} \rangle$, as a function of AGN X-ray
  luminosity at $z \approx 0$ (Left) and $z \approx 0.9$ (Right). The solid lines are
  the values predicted by Eq.~\ref{eq:avgM}, and the surrounding hatched areas are the 95\%
  confidence regions. The luminosity dependence in both quantities
  follows from the form of the total AGN CLFs derived in Paper I. At
  both redshifts, there is marginal evidence that Type 1 AGNs inhabit
  more massive haloes than Type 2 AGNs, in particular at $\log (L/\mathrm{erg\ s^{-1}}) \ga
  43$--$44$. However, the fact that the confidence regions overlap
  indicates that the difference in halo mass is not significant. The
  halo masses for $z \approx 0$ Type 1 quasars are likely
  overestimated due to limitations with the available data when
  deriving the total CLF (Paper I). The $z\approx 0$ Type 2 confidence
  region is substantially larger than the one for Type 1 AGNs because
  of the unconstrained upper-limit on $M_{\ast}$ (Table~\ref{table:res}).}
\label{fig:mass}
\end{figure*}

\begin{figure*}
\centering
\includegraphics[angle=-90,width=0.48\textwidth]{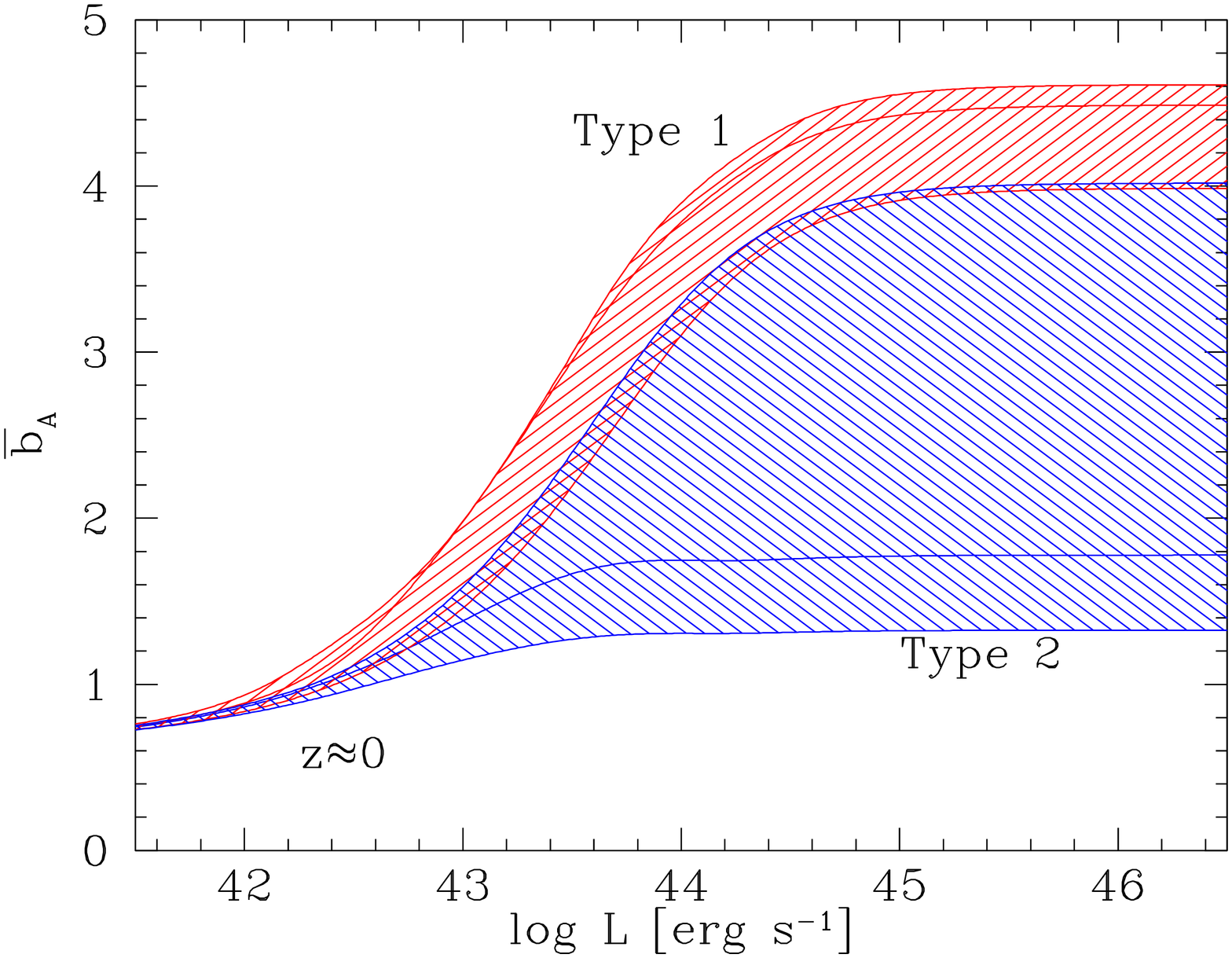}
\includegraphics[angle=-90,width=0.48\textwidth]{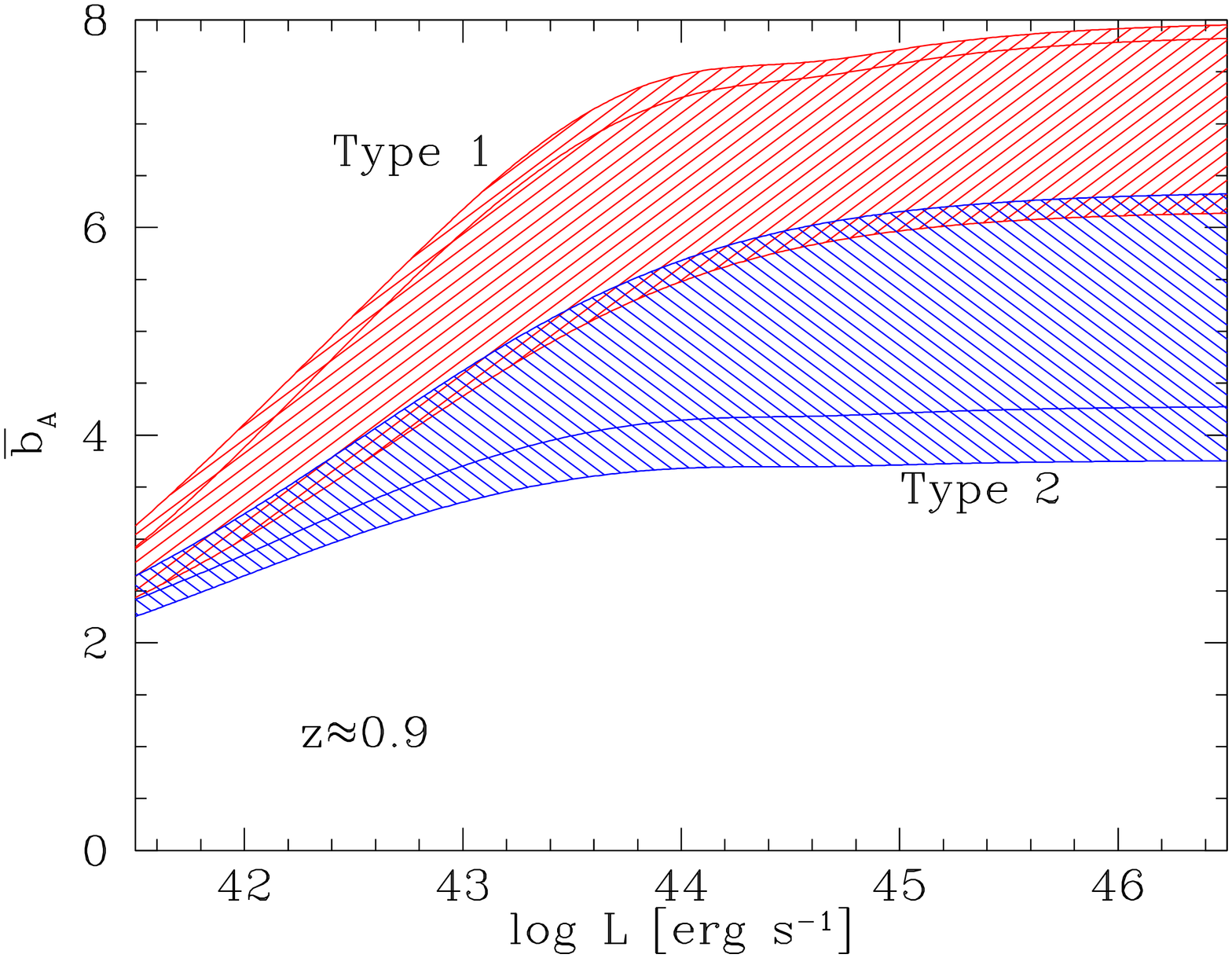}
\caption{The CLF-derived Type 2 and Type 1 mean AGN bias,
$\bar{b}_A$, as a function of AGN X-ray
  luminosity at $z \approx 0$ (Left) and $z \approx 0.9$ (Right). The solid lines are
  the values predicted by Eq.~\ref{eq:bbar}, and the surrounding hatched areas are the 95\%
  confidence regions. As with $\langle M_{\mathrm{h}} \rangle$, the luminosity dependence in both quantities
  follows from the form of the total AGN CLFs derived in Paper I. Also
  similar to the mean halo mass, the confidence regions of $\bar{b}_A$
  overlap at all luminosities indicating that any difference in
  average bias is not significant. As in Fig.~\ref{fig:mass}, the
  undetermined upper-limit to $M_{\ast}$ leads to a larger
  uncertainty in the predicted bias of Type~2 AGNs at $z\approx 0$.}
\label{fig:bias}
\end{figure*}

To see this explicitly, Eqs.~\ref{eq:bbar} and~\ref{eq:avgM} are used
to compute the average AGN bias, $\bar{b}_A$, and  mean
halo mass as a function of AGN luminosity, $\langle M_{\mathrm{h}}
\rangle(L)$, directly from the Type 1 and Type 2 CLFs at both $z$. The
resulting values, and their corresponding 95\% confidence regions are
plotted in Figs.~\ref{fig:mass} and~\ref{fig:bias}.
All of the $\langle M_{\mathrm{h}}\rangle(L)$ and $\bar{b}_A$ curves exhibit moderate
to strong luminosity dependence with larger values expected at
$\log (L/\mathrm{erg\ s^{-1}}) \ga 44$. This behaviour is a direct result of the total AGN
CLF derived from Paper I that was used in the calculation, and shows
that high and low luminosity AGNs of both types likely inhabit
different environments. Paper I discusses that this evolution in
luminosity is a strong indication that the triggering physics is
different in the two luminosity regimes.

In this paper, our focus is on any differences between the Type 2 and
Type 1 AGNs. Figs.~\ref{fig:mass} and~\ref{fig:bias} show that the confidence regions of
the two types overlap at all luminosities for both $z$. Therefore, as
expected from the $f_2(M_{\mathrm{h}})$ results, there is no
significant difference in the mean halo mass or bias between the two
populations. The predictions of the CLF model (solid lines) appear to
indicate a separation between the populations, in particular at higher
luminosities. This is marginal evidence that Type 1, unobscured
quasars, are more biased and inhabit more massive haloes than
obscured, Type 2, quasars. Interestingly, the sense of the difference in
halo mass agrees with the other X-ray based results \citep{allev11,allev14} at much
higher $z$. Lower-luminosity Seyfert galaxies, in contrast,
appear more likely to have the same $\bar{b}_A$ and $\langle M_{\mathrm{h}}
\rangle$ for both Type 2 and Type 1 AGNs, as expected for the
orientation-based model. Indeed, it is these types of AGNs in which
observations can directly probe the nature of any torus-like structure \citep[e.g.,][]{burt13}.

\section{Discussion and Summary}
\label{sect:discuss}
The question of whether or not the orientation-based unification model
is valid at all AGN luminosities and redshifts is important for 
constructing the correct model of black hole growth and galaxy
evolution. A potentially clear approach to this problem that avoids
interpreting the complex radiative environment around the AGN is to measure
the halo masses of the host galaxies of both obscured and unobscured
AGNs. If the orientation-based unification model holds the the host
haloes should be independent of the nuclear obscuration properties.

Building on earlier work on the clustering of galaxies \citep[e.g.,][]{yang03}, Paper I
presented a method to measure the AGN CLF at a particular
redshift. With the CLF several statistics of the host haloes of
AGNs can be computed as a function of luminosity, including the bias,
mass, and AGN occupation numbers. In this paper, we showed how the
total AGN CLF can be decomposed into ones describing Type 1 and Type
2 AGNs, and therefore the mean halo masses of both populations can be
computed as a function of luminosity at specific redshifts. Using
measurements of the Type 1 XLF by \citet{has05} and estimates of the
luminosity dependent Type 2 fraction $f_2(L)$ \citep{burlon11,merloni14}, the Type 1 and 2
CLFs were constrained at $z \approx 0$ and $0.9$. However, there was
no statistically significant difference in the mean
halo mass, $\langle M_{\mathrm{h}}\rangle(L)$, between Type 1 and 2
AGNs at either $z$ (Fig.~\ref{fig:mass}). The uncertainty in the predicted $\langle
M_{\mathrm{h}}\rangle(L)$ is significant, so the addition of further
data is needed to more precisely test if Type 1 and 2 AGNs inhabit
haloes of similar masses at all luminosities. The most useful data for
this purpose would be measurements of the correlation length at
different AGN luminosities, $r_0(L)$, for at least one of the AGN
types. In the X-ray band, such data may have to await for the surveys
performed by \textit{eROSITA}. 

Although the results are not significant, it is interesting to
consider the interpretation of the prediction shown in
Fig.~\ref{fig:mass}; namely, that the host galaxies of Type 1 AGNs
inhabit more massive haloes than Type 2 AGN host galaxies,
particularly at $\log (L/\mathrm{erg\ s^{-1}}) \ga 43$. As mentioned in
Sect.~\ref{sect:intro}, an alternative to the orientation-based
unification model is one where AGNs evolve from an obscured phase to
an unobscured phase. In this scenario, unobscured Type 1 quasars observed
at, e.g., $z \approx 0.9$ were triggered earlier and are at a
different point in their evolution than the obscured
Type 2 quasars observed at the same redshift. As this Type 1 phase
lasts longer than the obscured phase \citep[e.g.,][]{hopk08} then Type
1 quasars will be more frequently observed at a specific
$z$. Furthermore, as the Type 1s were triggered earlier than the Type 2s,
the dark matter halo hosting the Type 1 AGN will have accreted additional
mass prior to being observed. Thus, even if quasars are triggered in
haloes with a narrow range of masses \citep[e.g.,][]{hickox09}, the evolutionary model
predicts that obscured and unobscured AGNs will appear to reside in
different haloes when compared at the same $z$. For example, using
Eqn.~2 from \citet{fakh10} a $10^{14}$~M$_{\odot}$ halo at $z \sim 1$
will accrete $\approx 2\times 10^{13}$~M$_{\odot}$ in the $\sim$Gyr
timescale of the fading Type 1 quasar \citep{hopk06}. Such a change
would be challenging to detect at $z \la 1$, but might be possible at
higher redshifts when the halo accretion rates are larger \citep[cf.,][]{allev11,allev14}. Our CLF
results are consistent with this small change, but larger differences
are possible which, if confirmed by revising the CLF fits with
improved data, would lead to a substantial revision to our
understanding of AGN physics.  

\begin{figure}
\centering
\includegraphics[angle=-90,width=0.48\textwidth]{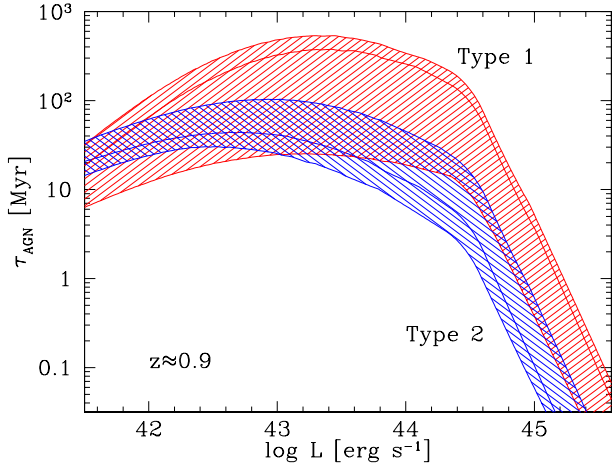}
\caption{The CLF-derived Type 2 and Type 1 AGN lifetimes,
$\tau_{\mathrm{AGN}}$, as a function of AGN X-ray
  luminosity at $z \approx 0.9$. The solid lines are
  the values predicted by Eq.~\ref{eq:time}, and the surrounding hatched areas are the 95\%
  confidence regions. Although, again, the results are not
  significant, there are indications that at $\log (L/\mathrm{erg\ s^{-1}}) \ga 44$ the lifetimes of Type 1 AGNs
  are longer than Type 2s and can last several hundreds of Myrs, consistent with the evolutionary picture of
  AGNs.}
\label{fig:time}
\end{figure}

As a check on these ideas, Fig.~\ref{fig:time} plots the estimated
AGN Type 1 and Type 2 lifetimes at $z \approx 0.9$ computed from the CLFs (via
Eq.~\ref{eq:time}).
These lifetimes are estimated from the ratio of the AGN space density
to the dark matter halo density \citep{mw01}, and are therefore
difficult to directly compare to the predictions of, for example,
merger-triggered AGN evolution where the luminosity of an AGN varies
significantly over its lifetime \citep[e.g.,][]{hopk08}. Nevertheless,
this snapshot of the AGN population at $z \approx 0.9$ shows an
interesting evolution across luminosity that can be explored in future
work. The figure shows that, although the confidence regions formally
overlap at all luminosities, the overlap is smallest at $\log (L/\mathrm{erg\ s^{-1}}) \ga
44$. Below that luminosity, Type 1 and 2 AGNs appear to have similar
lifetimes, but above $\log (L/\mathrm{erg\ s^{-1}}) \sim 44$ the Type 1 lifetime may be
larger, and last hundreds of Myrs, consistent with the evolutionary model described above. 

Observationally, differences in the host halo masses of Type 1 and 2
AGNs have been typically been very small \citep{dhm15}, or non-existent
\citep{mendez16}. Interestingly, the studies that have found that unobscured AGNs
have larger halo masses than Type 2 objects use exclusively X-ray
selected samples \citep{allev11,allev14}. Analyses that find the opposite trend
(haloes hosting obscured AGN are more massive) often utilize IR
selected AGN samples. As emphasized by \citet{mendez16}, the different
wavelengths select different types of host galaxies, and, since the
size of the true effect may be small, its not clear if the two sets of
measurements represent a fundamental disagreement. For example, a
substantial fraction of the infrared AGNs are so cocooned by gas and dust that they
would are undetected in X-ray surveys
\citep[e.g.,][]{don14}. Therefore, studies based on these samples may
be selecting a specific time in AGN evolution. The CLF-derived
results presented here make use of X-ray selected AGNs, and the trend
we uncover is consistent with the previously mentioned X-ray results. 

Finally, it is worth noting that the method described in
Sect.~\ref{sect:CLF} can be applied to any subset of AGNs with which
there are appropriate datasets. For example, individual CLFs for
radio-loud and radio-quiet AGNs, or Compton-thick and Compton-thin AGNs could be
derived using this procedure. Data produced from future surveys by 
\textit{eROSITA}, LSST, \textit{Euclid}, \textit{WFIRST} and \textit{Athena} can then
be exploited to provide a comprehensive view of multiple facets of the
AGN phenomenon and how it connects to the overall cosmic web.

\section*{Acknowledgments}
The author thanks J.\ Wise, G.\ Altay, A.\ Myers, and R.\ Grissom for
help and advice during the course of this work, R.\ Hickox for
comments on a draft of the manuscript, and acknowledges
support from NSF award AST 1333360.

%%%%%%%%%%%%%%%%%%%%%%%%%%%%%%%%%%%%%%%%%%%%%%%%%%



%\bibliographystyle{mnras}

\begin{thebibliography}{99}
\bibitem[\protect\citeauthoryear{Allevato \etal}{2011}]{allev11}
Allevato V., \etal, 2011, \apj, 736, 99
\bibitem[\protect\citeauthoryear{Allevato \etal}{2014}]{allev14}
Allevato V., \etal, 2014, \apj, 796, 4
\bibitem[\protect\citeauthoryear{Antonucci}{1993}]{anton93} Antonucci
  R., 1993, \araa, 31, 473
\bibitem[\protect\citeauthoryear{Ballantyne}{2014}]{ball14}
    Ballantyne D.R., 2014, \mnras, 437, 2845 
\bibitem[\protect\citeauthoryear{Ballantyne}{2016}]{ball16} Ballantyne
  D.R., 2016, \mnras, submitted (Paper I)
\bibitem[\protect\citeauthoryear{Ballantyne, Everett \&
    Murray}{2006}]{bem06} Ballantyne D.R., Everett J.E., Murray N.,
  2006, \apj, 639, 740
\bibitem[\protect\citeauthoryear{Buchner \etal}{2015}]{buch15} Buchner J.,
  Georgakakis A., Nandra K., Brightman M., Menzel M.-L., Liu Z., Hsu
  L-T., Salvato M., Rangel C., Aird J., Merloni A., Ross N., 2015,
  \apj, 802, 89
\bibitem[\protect\citeauthoryear{Burlon \etal}{2011}]{burlon11} Burlon
D., Ajello M., Greiner J., Comastri A., Merloni A., Gehrels N., 2011,
\apj, 728, 58
\bibitem[\protect\citeauthoryear{Burtscher \etal}{2013}]{burt13}
  Burtscher L., \etal, 2013, \aap, 558, 149
\bibitem[\protect\citeauthoryear{Burtscher \etal}{2016}]{burt16}
  Burtscher L., \etal, 2016, \aap, 586, A28
\bibitem[\protect\citeauthoryear{Cappelluti \etal}{2010}]{cap10}
  Cappelluti N., Ajello M., Burlon D., Krumpe M., Miyaji T.,
  Bonoli S., Griener, J., 2010, \apj, 716, L209
\bibitem[\protect\citeauthoryear{Cappelluti \etal}{2012}]{caf12}
  Cappelluti N., Allevato V., Finoguenov, A., 2012, Adv. in Astron.,
  853701
\bibitem[\protect\citeauthoryear{Del Moro \etal}{2016}]{delmoro16} Del
  Moro A., \etal,  2016, \mnras, 456, 2105
\bibitem[\protect\citeauthoryear{Di Matteo \etal}{2005}]{dsh05} Di
  Matteo T., Springel V.,  Hernquist L., 2005, \nat, 433, 604
\bibitem[\protect\citeauthoryear{DiPompeo \etal}{2014}]{dipomp14}
  DiPompeo M.A., Myers A.D., Hickox R.C., Geach J.E., Hainline K.N.,
  2014, \mnras, 442, 3443
\bibitem[\protect\citeauthoryear{DiPompeo, Hickox \&
    Myers}{DiPompeo \etal}{2016}]{dhm15} DiPompeo M.A., Hickox R.C., Myers A.D., 2016,
  \mnras, 456, 924
\bibitem[\protect\citeauthoryear{Donoso \etal}{2014}]{don14} Donoso
  E., Yan L., Stern D., Assef R., 2014, \apj, 789, 44
\bibitem[\protect\citeauthoryear{Draper \& Ballantyne}{2012}]{db12}
  Draper A.R., Ballantyne D.R., 2012, \apj, 751, 72
\bibitem[\protect\citeauthoryear{Fakhouri, Ma \&
    Boylan-Kolchin}{2010}]{fakh10} Fakhouri O., Ma C.-P.,
  Boylan-Kolchin M., 2010, \mnras, 406, 2267
\bibitem[\protect\citeauthoryear{Geach \etal}{2013}]{geach13} Geach
  J.E., \etal, 2013, \apj, 776, L41
\bibitem[\protect\citeauthoryear{Glikman \etal}{2015}]{glik15} Glikman
  E., Simmons B., Mailly M., Schawinski K., Urry C.M., Lacy, M., 2015,
  \apj, 806, 218
\bibitem[\protect\citeauthoryear{Hasinger}{2008}]{has08} Hasinger G.,
  2008, \aap, 490, 905
\bibitem[\protect\citeauthoryear{Hasinger, Miyaji \& Schmidt}{Hasinger
    \etal}{2005}]{has05} Hasinger G., Miyaji T., Schmidt M., 2005,
  \aap, 441, 417
\bibitem[\protect\citeauthoryear{Hernquist}{1989}]{hernq89} Hernquist
  L., 1989, \nat, 340, 687
\bibitem[\protect\citeauthoryear{Hickox \etal}{2009}]{hickox09} Hickox
  R.C., \etal, 2009, \apj, 696, 891
\bibitem[\protect\citeauthoryear{Hickox \etal}{2011}]{hick11} Hickox
  R.C., \etal, 2011, \apj, 731, 117
\bibitem[\protect\citeauthoryear{Hinshaw \etal}{2013}]{hinshaw13}
  Hinshaw G., \etal, 2013, \apjs, 208, 19
\bibitem[\protect\citeauthoryear{H\"{o}nig \etal}{2013}]{honig13}
  H\"{o}nig S.F., \etal, 2013, \apj, 771, 87
\bibitem[\protect\citeauthoryear{Hopkins \etal}{2005}]{hopk05}
  Hopkins P.F., Hernquist L., Cox T.J., Di Matteo T., Martini P.,
  Robertson B., Springel V., 2005, \apj, 630, 705
\bibitem[\protect\citeauthoryear{Hopkins \etal}{2006}]{hopk06} Hopkins
  P.F., Hernquist L., Cox T.J., Di Matteo T., Robertson B., Springel
  V., 2006, \apjs, 163, 1
\bibitem[\protect\citeauthoryear{Hopkins \etal}{2008}]{hopk08}
  Hopkins P.F., Hernquist L., Cox T.J., Kere\v{s}, D.,
  2008, \apjs, 175, 356
\bibitem[\protect\citeauthoryear{Hopkins \etal}{2014}]{hkb14}
  Hopkins P.F., Kocevski D.D., Bundy K., 2014, \mnras, 445, 823
\bibitem[\protect\citeauthoryear{Kauffmann \& Haehnelt}{2000}]{kh00}
  Kauffmann G., Haehnelt M., 2000, MNRAS, 311, 576
\bibitem[\protect\citeauthoryear{Kocevski \etal}{2012}]{koc12}
  Kocevski D., \etal, 2012, \apj, 744, 148
%\bibitem[\protect\citeauthoryear{Kolodzig \etal}{2013a}]{kold13}
%  Kolodzig A., Gilfanov M., Sunyaev R., Sazonov S., Brusa, M.,
%  2013, \aap, 558, A89
%\bibitem[\protect\citeauthoryear{Kolodzig \etal}{2013b}]{kold13b}
%  Kolodzig A., Gilfanov M., H\"{ut}tsi G., Sunyaev, R., 2013, \aap,
%  558, A90
\bibitem[\protect\citeauthoryear{Koutoulidis \etal}{2013}]{kou13}
  Koutoulidis L., Plionis M., Georgantopoulos I., Fanidakis, N.,
  2013, \mnras, 428, 1382
\bibitem[\protect\citeauthoryear{La Franca \etal}{2005}]{laf05} La
  Franca F., \etal, 2005, \apj, 635, 864
\bibitem[\protect\citeauthoryear{Leauthaud \etal}{2015}]{lea15}
  Leauthaud A., \etal, 2015, \mnras, 446, 1874
\bibitem[\protect\citeauthoryear{Martini \& Weinberg}{2001}]{mw01}
  Martini P., Weinberg D.H., 2001, \apj, 547, 12
\bibitem[\protect\citeauthoryear{Menci \etal}{2006}]{menci06} Menci
  N., Fontana A., Giallongo E., Grazian A., Salimbeni S.,
  2006, \apj, 647, 753
\bibitem[\protect\citeauthoryear{Mendez \etal}{2016}]{mendez16} Mendez
  A.J., \etal, 2016, \apj, 821, 55
\bibitem[\protect\citeauthoryear{Merloni \etal}{2014}]{merloni14}
  Merloni A., \etal, 2014, \mnras, 437, 3550
\bibitem[\protect\citeauthoryear{Metropolis \etal}{1953}]{metro53}
  Metropolis N., Rosenbluth A.W., Rosenbluth M.N., Teller A.H.,
  Teller, E., 1953, J. Chem. Phys., 21, 1087
\bibitem[\protect\citeauthoryear{Netzer}{2015}]{netzer15} Netzer H.,
  2015, \araa, 53, 365
\bibitem[\protect\citeauthoryear{Sanders \etal}{1988}]{san88} Sanders
  D.B., Soifer B.T., Elias J.H., Madore B.F., Matthews K.,
  Neugebauer G., Scoville N.Z., 1988, \apj, 325, 74
\bibitem[\protect\citeauthoryear{Sazonov, Churazov \&
    Krivonos}{Sazonov \etal}{2015}]{sck15} Sazonov S., Churazov E.,
  Krivonos R., 2015, \mnras, 454, 1202
\bibitem[\protect\citeauthoryear{Springel, Di Matteo \& Hernquist}{2005}]{sdh05}
  Springel V., Di Matteo T., Hernquist L., 2005, \mnras, 361, 776
\bibitem[\protect\citeauthoryear{Tinker \etal}{2010}]{tink10} Tinker
  J.L., Robertson B.E., Kravtsov A.V., Klypin A., Warren M.S., Yepes
  G., Gottl\"{o}ber S., 2010, \apj, 724, 878
\bibitem[\protect\citeauthoryear{Treister \& Urry}{2006}]{tu06}
  Treister E., Urry C.M., 2006, \apj, 652, L79
\bibitem[\protect\citeauthoryear{Treister \etal}{2012}]{trei12}
  Treister E., Schawinski K., Urry C.M., Simmons B.D., 2012, \apj,
  758, L39
\bibitem[\protect\citeauthoryear{Tristram \etal}{2014}]{trist14}
  Tristram K.R.W., Burtscher L., Jaffe W., Meisenheimer K., H\"{o}nig
  S.F., Kishimoto M., Schartmann M., Weigelt G., 2014, \aap, 563, 82
\bibitem[\protect\citeauthoryear{Ueda \etal}{2014}]{ueda14} Ueda Y.,
  Akiyama M., Hasinger G., Miyaji T., Watson M.G.,
  2014, \apj, 786, 104
\bibitem[\protect\citeauthoryear{Urry \& Padovani}{1995}]{up95} Urry
  C.M., Padovani P., 1995, \pasp, 107, 803
\bibitem[\protect\citeauthoryear{van den Bosch \etal}{2003}]{vym03}
  van den Bosch F.C., Yang X., Mo H.J., 2003, \mnras, 340, 771
\bibitem[\protect\citeauthoryear{van den Bosch \etal}{2007}]{van07}
  van den Bosch F., \etal, 2007, \mnras, 376, 841
\bibitem[\protect\citeauthoryear{Yang \etal}{2003}]{yang03} Yang X.,
  Mo H.J., van den Bosch, F., 2003, \mnras, 339, 1057
\end{thebibliography}


%%%%%%%%%%%%%%%%%%%%%%%%%%%%%%%%%%%%%%%%%%%%%%%%%%

%%%%%%%%%%%%%%%%% APPENDICES %%%%%%%%%%%%%%%%%%%%%

\appendix
\section{A Check on Internal Self-Consistency}
\label{app:self}
A useful check on the measured CLF is to compute the
luminosity-dependence of the mean Type~2 fraction, $\bar{f}_2(L)$, directly from the CLF
using Eq.~\ref{eq:ty2vsL}. Specific forms for $\bar{f}_2(L)$ are assumed
during the computation of $f_2(L,M_{\mathrm{h}})$ (i.e.,
Eq.~\ref{eq:f2burlon} or~\ref{eq:f2merloni}), but during the fit
$f_2(L,M_{\mathrm{h}})$ is also subject to the requirement that it is
bounded by unity at all $L$ and \mh. Therefore, computing $\bar{f}_2(L)$
directly from Eq.~\ref{eq:ty2vsL} following the fit will provide a
check on the self-consistency of the solution, as well as our
assumption that $f_2(L,M_{\mathrm{h}})$ is separable into individual
functions of $L$ and $M_{\mathrm{h}}$.

\begin{figure*}
\includegraphics[angle=-90,width=0.48\textwidth]{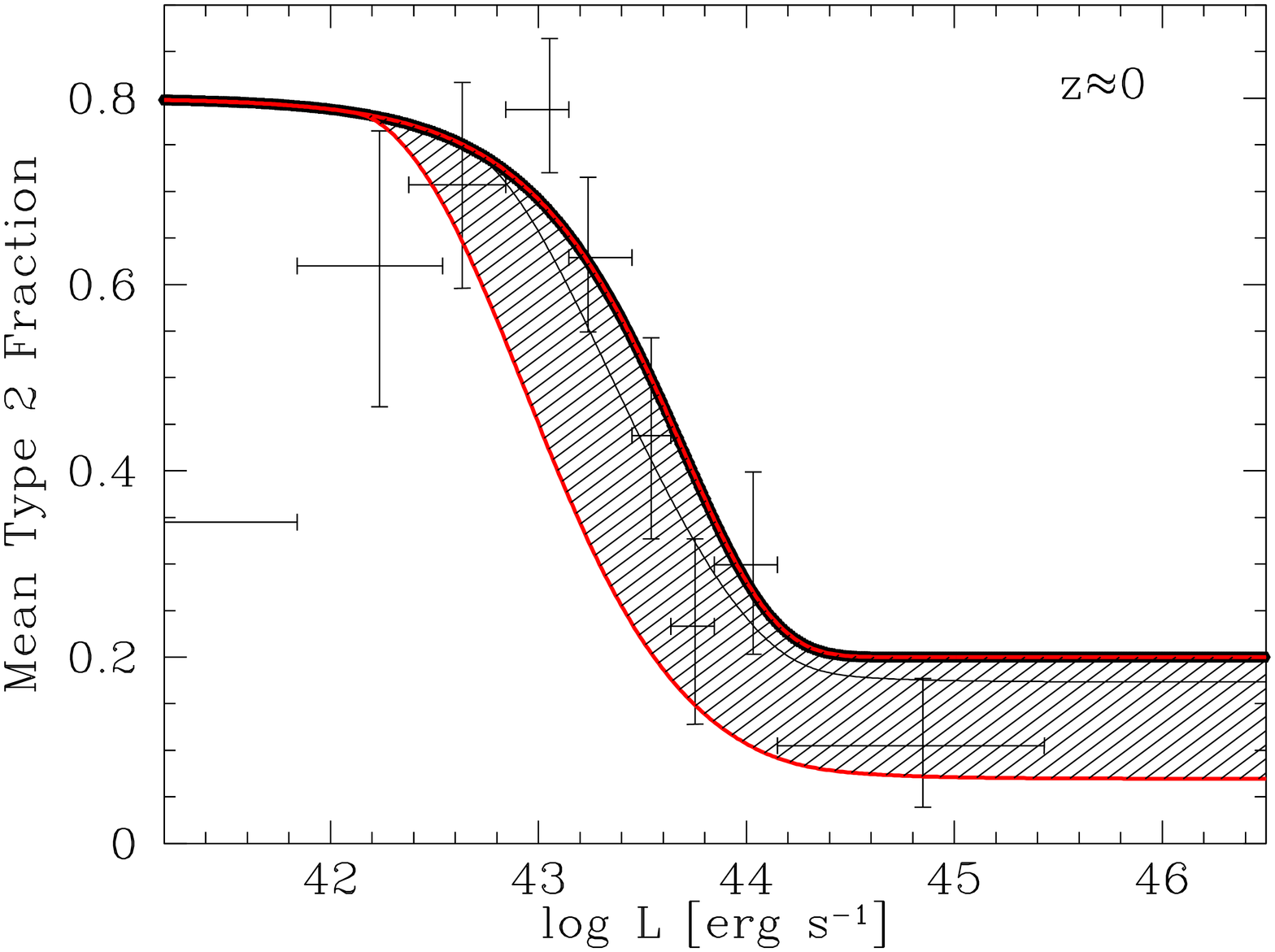}
\includegraphics[angle=-90,width=0.48\textwidth]{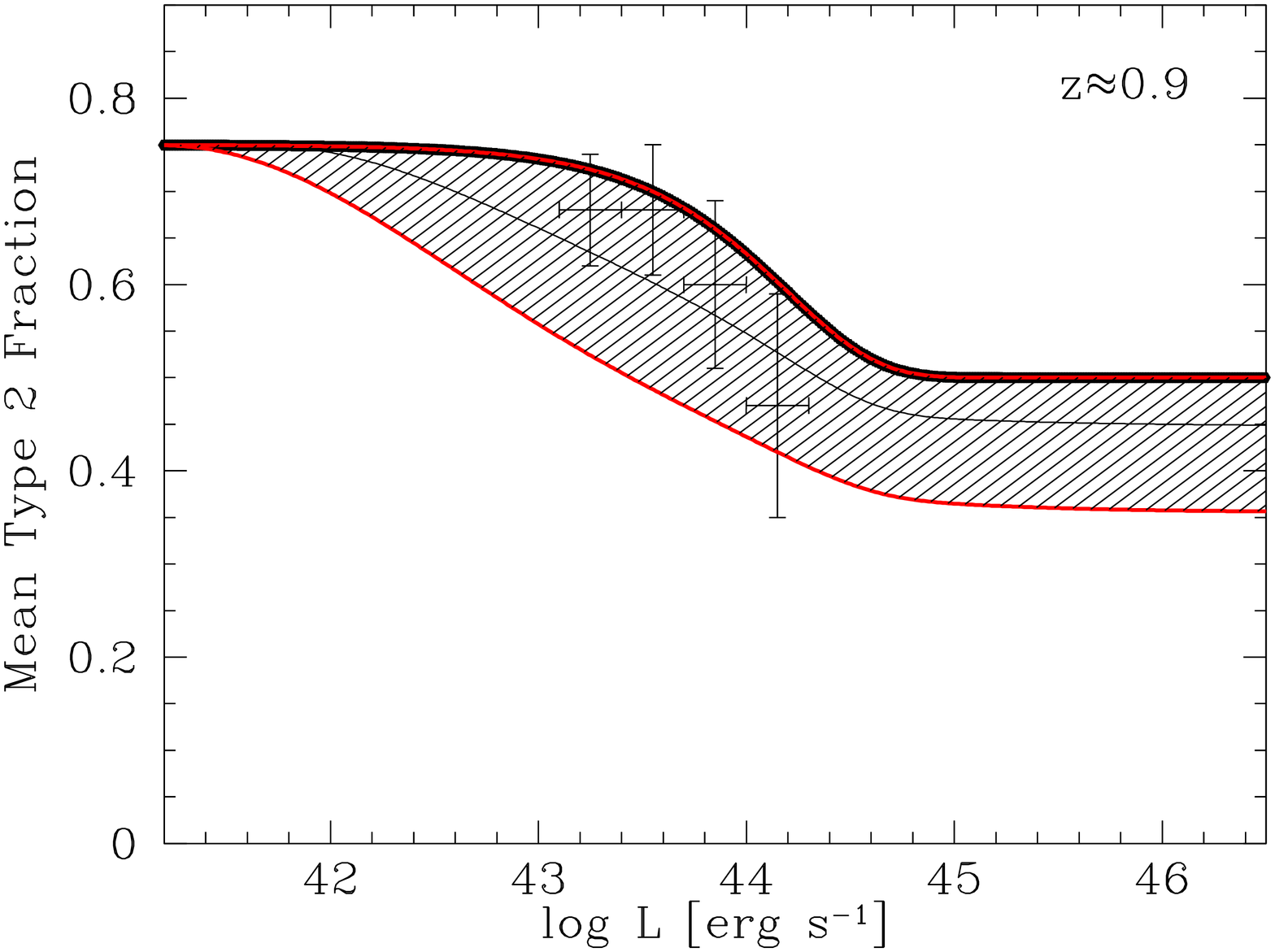}
\caption{(Left) The mean Type 2 fraction versus AGN luminosity at $z
  \approx 0$. The thick black line plots the relation (Eq.~\ref{eq:f2burlon})
  from \citet{burlon11} that was used as an input into the Type 2 and
  Type 1 CLF calculation (Sect.~\ref{sub:CLFhowto}). The thin solid line
  and the hatched region plot the mean Type 2 fraction (and its 95\%
  confidence region, bounded by the red lines) calculated from the
  best fit CLF model after fitting the data
  (Eq.~\ref{eq:ty2vsL}). The data points show the \bat\ derived
  Type 2 fractions measured by \citet{burlon11}. The mean Type 2 fractions
  calculated by the CLF model remains consistent with the
  \citet{burlon11} data and model. (Right) As in the left-hand panel,
  but now at $z \approx 0.9$. The thick black line plots
  Eq.~\ref{eq:f2merloni} and the data points are the Type 2 fractions
  measured by \citet{merloni14} at $0.8 \leq z \leq 1.1$. Again, the
  CLF-derived mean Type 2 fractions are consistent with the input assumptions.}
\label{fig:ty2vsL}
\end{figure*}

Both panels of Figure~\ref{fig:ty2vsL} compares $\bar{f}_2(L)$ computed from
Eq.~\ref{eq:ty2vsL} directly (thin black solid lines) with the one input
into the calculation (thick black lines).
The hatched regions (bounded by the red lines) show the $95$\% confidence regions on the
CLF-derived $\bar{f}_2(L)$, and the data points are the measurements
on which the Eqs.~\ref{eq:f2burlon} and~\ref{eq:f2merloni} are
based. The figure shows that at both redshifts the $\bar{f}_2(L)$
computed from the CLF model is consistent with both the observed Type
2 fractions and the input assumptions. More precise measurements of
$f_2(L)$ will be important to improve the modeling of the Type 1 and
Type 2 AGN CLFs.

%%%%%%%%%%%%%%%%%%%%%%%%%%%%%%%%%%%%%%%%%%%%%%%%%%

% Don't change these lines
\bsp	% typesetting comment
\label{lastpage}
\end{document}